\documentclass[reprint,amsmath,amssymb,notitlepage,citeautoscript,nonbibnotes,nofootinbib]{revtex4-1}

\usepackage{bm,mathrsfs,dcolumn,graphicx,color}
\usepackage{amsmath}
\usepackage{graphicx}
\usepackage[T1]{fontenc}
\usepackage{hyphenat}
\usepackage{setspace}
\usepackage{footnote}
\usepackage[colorlinks]{hyperref}

\begin{document}

\title{{\it Moscow University Physics Bulletin {\bf 3}, 30 (1967)} \\
$\mbox{   }$\\
Screened potential of a point charge in a thin film}

\author{N. S. Rytova}\thanks{Translated from the original in Russian by Alexey Chernikov \& Mikhail M. Glazov (2017).\\ Original Russian text is available here: \href{http://vmu.phys.msu.ru/abstract/1967/3/1967-3-030/}{http://vmu.phys.msu.ru/abstract/1967/3/1967-3-030/}\\
Original English translation is available here: \href{http://vmu.phys.msu.ru/abstract/1967/3/1967-3-030/}{http://vmu.phys.msu.ru/abstract/1967/3/1967-3-030/}\\
Direct link: \href{https://www.dropbox.com/s/fe1eax2li8udnpj/en-67-22-3-18.pdf?dl=0}{https://www.dropbox.com/s/fe1eax2li8udnpj/en-67-22-3-18.pdf}}
\affiliation{Department of Semiconductor Physics, Moscow State University, Moscow, USSR}

\begin{abstract}

The potential of a point charge in a thin semiconducting film with the thickness below the \mbox{de Broglie} wavelength of the free charge carriers is calculated with and without the screening [\textit{by free charge carriers}]\footnote{Hereafter the italic text in square brackets is added in translation.}.
\end{abstract}
\maketitle

For a number of problems in thin film physics, it is necessary to know the interaction energy between point charges, i.e., the form of the potential created by a point charge in a thin film. For conducting films, it is further important to obtain the law for the screening of the point charge by the free electrons. 

Spatial inhomogeneity of the system leads to deviations both for an unscreened potential of a point charge and for the screening law in comparison to a bulk crystal. In the limit of a very thin film, both have a specific, ``two-dimensional'' character. Here, we will obtain expressions for the screened and unscreened potentials in this limiting case and present the criteria for the validity of the approximation.

Let us consider a thin film with a thickness $L$ being smaller than the de Broglie wavelength of the free electrons, i.e., where the following relations hold: $L\ll1/\sqrt{N}$ in the degenerate case and $L\ll\hbar/\sqrt{2mk_{B}T}$ for a Boltzmann distribution. Here, $N$ is the electron surface density, $T$ -- the temperature, and $k_{B}$ -- the Boltzmann constant. (For a semiconducting film with the thickness $L=100$\,\AA~and the carrier effective mass $m=0.1\,m_0$, these relations are valid for [\textit{bulk}] carrier concentrations $n$\,<$10^{18}$\,cm$^{-3}$ and temperatures $T<30$\,K.) Then, the motion of the electron along the normal of the film plane is quantized and only the lowest subband is effectively occupied for the simultaneous validity of the above two relations to hold. Under these conditions, the average distance between free electrons and the immobile [\textit{positive}] point charges, responsible for the total charge neutrality, is much larger than the film thickness. Hence, it is particularly important to determine the interaction potential specifically at these distances.

Therefore, the aim is to find the asymptotic behavior of the unscreened and screened potential of a point charge located in a thin film for distances larger than the film thickness.

\section*{Unscreened potential}

Let the dielectric constant of the environment be $\varepsilon_1$\footnote{\textit{i.e., $\varepsilon_1$=$\varepsilon_3$}}, that of the film be $\varepsilon_2$, and their ratio be denoted as $\varepsilon =  \varepsilon_2/\varepsilon_1$. The $z$-axis is normal to the plane of the film (see Fig. 1) and we consider the film to be infinite in the $xy$-plane 

\begin{figure}[ht]
\includegraphics[width=4 cm]{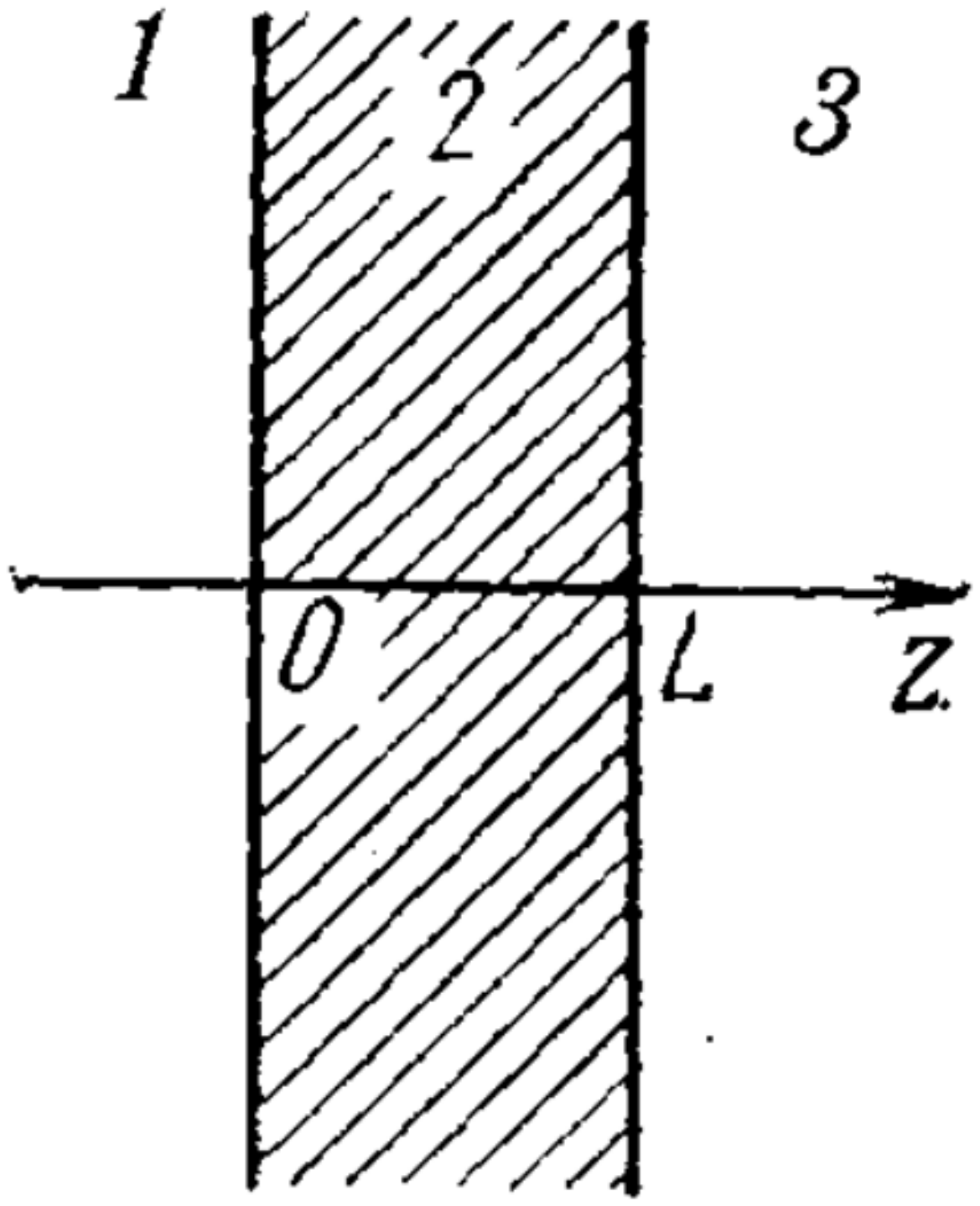}
\caption{\label{fig1} [\textit{Schematic illustration of the system geometry.}]
}
\end{figure}

The potential $\varphi(\vec{r},\vec{r}\,')$ at the position $\vec{r}$, created by the point charge located at $\vec{r}\,'$, satisfies the following equations in the regions 1, 2, and 3:
\begin{align}
\label{eq1}
\nabla^2_{\vec{r}}\,\varphi_{1}(\vec{r},\vec{r}\,')&=0,\nonumber\\
\nabla^2_{\vec{r}}\,\varphi_{2}(\vec{r},\vec{r}\,')&=-\frac{4\pi e}{\varepsilon_2}\delta(\vec{r}-\vec{r}\,'),\\
\nabla^2_{\vec{r}}\,\varphi_{3}(\vec{r},\vec{r}\,')&=0\nonumber
\end{align}
with the boundary conditions
\begin{align}
\label{eq2}
\text{at}~z=0,~\varphi_1=\varphi_2,~\frac{\partial \varphi_1}{\partial z}=\varepsilon\frac{\partial \varphi_2}{\partial z}\nonumber\\
\text{at}~z=L,~\varphi_2=\varphi_3,~\varepsilon\frac{\partial \varphi_2}{\partial z}=\frac{\partial \varphi_3}{\partial z}
\end{align} 
and the requirement to be bounded at the infinity. 

Due to the homogeneity and isotropy of the system in the $xy$-plane of the film, $\varphi(\vec{r},\vec{r}\,')$ depends on $\left|\vec{\rho}-\vec{\rho}\,'\right|$, where $\vec{\rho}$ is the in-plane position vector. It allows us, without loss of generality, to set $\vec{\rho}\,'=0$, i.e., to place the point charge onto the $z$-axis and to expand the potential $\varphi(\vec{r},\vec{r}\,')$ into a double-Fourier-integral:
\begin{align}
\label{eq3}
\varphi(\vec{r},\vec{r}\,') = \int \frac{d\vec{k}}{(2\pi)^2} e^{i \vec{k}\vec{\rho}} \varphi(k,z,z').
\end{align} 

The Fourier components $\varphi(k, z, z\,')$ depend on the absolute value $|\vec{k}|$. Inserting \eqref{eq3} into \eqref{eq1} and \eqref{eq2}, we obtain the following equations for the Fourier components $\varphi(k, z, z\,')$ of the potential:
\begin{align}
&\frac{\partial^2\varphi_1(k,z,z')}{\partial z^2} - k^2 \varphi_1(k,z,z')=0, \nonumber\\
&\frac{\partial^2\varphi_2(k,z,z')}{\partial z^2} - k^2 \varphi_2(k,z,z')= - \frac{4\pi e}{\varepsilon_2} \delta(z-z'), \label{eq4}\\
&\frac{\partial^2\varphi_3(k,z,z')}{\partial z^2} - k^2 \varphi_3(k,z,z')=0,\nonumber
\end{align}
with the boundary conditions~\eqref{eq2}.

In the region of interest inside the film [\textit{region 2, see Fig.\,\ref{fig1}}], the solution of the Eqs.~\eqref{eq4} has the following form:
\begin{multline}
\label{eq5}
\varphi_2(k,z,z') = \frac{2\pi e}{\varepsilon_2 k} \left\{ e^{-k|z-z'|} \right. + \\ \left.
\frac{2\delta}{e^{2kL} - \delta^2} [\delta \cosh{k(z-z')} + e^{kL} \cosh{k(z+z'-L)}] \right\}
\end{multline}
where
\begin{align}
\delta=\frac{\varepsilon-1}{\varepsilon+1}. \nonumber
\end{align}
In order to obtain the asymptotic behavior for the potential $\varphi_2(\vec{r},\vec{r}\,')$ for distances $\rho \gg L$, it is sufficient to know  its Fourier component $\varphi_2(k, z, z\,')$ for $kL \ll 1$. Assuming $kL \ll 1$ in Eq.\,\eqref{eq5}, we obtain an expression, independent from $z$ and $z\,'$
\begin{equation}
\label{eq6}
\varphi(k) = \frac{2\pi e}{\varepsilon_2 k} \left[\frac{e^{kL}+\delta}{e^{kL}-\delta}\right].
\end{equation}
The exponents in \eqref{eq6} are not expanded into power series, since the absolute value of $\delta$ can be close to unity, in which case only one of the quantities $(1-\delta)$ or $(1+\delta)$ can be comparable to $kL$.

This way, for distances larger than the film thickness, the potential of the point charge does not depend on the coordinates $z$ and $z\,'$ and becomes ``two-dimensional''. Let us obtain the expressions for it in three limiting cases.

\textbf{1.} $\varepsilon \gg 1$, i.e., the optical density\footnote{Note added in translation: We keep the authors term ``optical density'' although the static screening is discussed throughout the paper.} of the film is larger than that of its surroundings. Then, $\delta=1-{2}/{\varepsilon}$ and the expression \eqref{eq6} assumes the form of
\begin{equation}
\label{eq7}
\varphi(k) = \frac{4\pi e}{\varepsilon_2 Lk\left(k + \frac{2}{\varepsilon L}\right)}.
\end{equation}
Inserting \eqref{eq7} into \eqref{eq3}, we find for $\rho \gg L$
\begin{align}
\label{eq8}
\varphi(\rho)&=\frac{2e}{\varepsilon_2 L}\int\limits_{0}^{\infty}{{dk}\frac{J_0(k\rho)}{k+\frac{2}{\varepsilon L}}}\nonumber\\
&=\frac{\pi e}{\varepsilon_2 L}\left[\mathbf H_0\left(\frac{2\rho}{\varepsilon L}\right)-N_0\left(\frac{2\rho}{\varepsilon L}\right)\right],
\end{align}
where $\mathbf H_0(x)$, $N_0(x)$, and $J_0(x)$ are the zero-order Struve, Neumann, and Bessel functions, respectively.

At very large distances $\rho \gg \varepsilon L/2$, the expression \eqref{eq8} takes the form
\begin{align}
\label{eq9}
\varphi(\rho)=\frac{e}{\varepsilon_1 \rho},
\end{align}
i.e., the potential very far away from the point charge is such, as if the film did not exist at all. It is the Coulomb potential in a homogeneous dielectric medium with the permittivity $\varepsilon_1$.

\textbf{2.} $\varepsilon \approx 1$, i.e., the optical density of the film and the environment are roughly equal. In this case, $|\delta|\ll 1$ and
\begin{align}
\label{eq10}
\varphi(k)=\frac{2\pi e}{\varepsilon_1 k}=\frac{2\pi e}{\varepsilon_2 k}.
\end{align}

Since $\varepsilon_1\approx\varepsilon_2$, the system becomes nearly homogeneous and by inserting \eqref{eq10} into \eqref{eq3}, for distances $\rho\gg L$, we obtain the Coulomb potential:
\begin{align}
\varphi(\rho)=\frac{e}{\varepsilon_2}\int\limits_{0}^{\infty}{J_0(k\rho)dk}=\frac{e}{\varepsilon_2 \rho}=\frac{e}{\varepsilon_1 \rho}.\nonumber
\end{align}

\textbf{3.} $\varepsilon \ll 1$, i.e., the optical density of the film is much smaller than that of the surroundings. Here, $\delta=-(1-2\varepsilon)$ and
\begin{align}
\varphi(k)=\frac{2\pi e \varepsilon}{\varepsilon_2 k}=\frac{2\pi e}{\varepsilon_1 k},\nonumber
\end{align}
i.e., also in this case, already for $\rho \gg L$, the potential is such, as if the film did not exist:
\begin{align}
\varphi(\rho)=\frac{e}{\varepsilon_1 \rho}.\nonumber
\end{align}

\section*{Screened potential}

Let the film contain free charges (electrons) distributed with an average surface density $N$ and immobile charges of the opposite sign with the same average surface density. Overall, the film is charge neutral and macroscopically homogenous in the $xy$-plane. 

Screened potential $\tilde{\varphi}(\vec{r},\vec{r}\,')$ of the point charge at the position $\vec{r}\,'$ of the film (as before, we set $\vec{\rho}\,'=0)$ satisfies in the region 2 the equation
\begin{align}
\label{eq11}
\nabla_{\vec{r}}^2 \,\tilde \varphi_2(\vec{r}, \vec{r}\,') = -\frac{4\pi e}{\varepsilon_2} \left[ \delta(\vec{r} - \vec{r}\,') - \Delta n(\vec{r},\vec{r}\,')\right],
\end{align}
where $\Delta n(\vec{r},\vec{r}\,')$ is the change of the free electron concentration in $\vec{r}$ under the influence of the field of a point charge located at $\vec{r}\,'$.\\
~\\
~~~~Under thermal equilibrium conditions, the electron concentration is a function of the electrochemical potential
\begin{align}
\mu=\mu_0+e\tilde{\varphi},\nonumber
\end{align}
where $\mu_0$ is the chemical potential of the electrons in the absence of a field. For sufficiently large distances, when $e\tilde{\varphi}\ll\mu_0$, it is possible to expand $\Delta n$ into a series in $\tilde{\varphi}$ and consider only the first linear term
\begin{align}
\label{eq12}
\Delta n = \frac{\partial n}{\partial \mu_0}e\tilde{\varphi}.
\end{align}
Inserting \eqref{eq12} into \eqref{eq11} and expanding $\tilde{\varphi}(\rho, z, z\,')$ into a Fourier integral as exemplified in \eqref{eq3}, we obtain the equations for the Fourier components
\begin{widetext}
\begin{align}
&\frac{\partial^2\tilde\varphi_1(k,z,z')}{\partial z^2} - k^2 \tilde\varphi_1(k,z,z')=0, \nonumber\\
&\frac{\partial^2\tilde\varphi_2(k,z,z')}{\partial z^2} - \left[ k^2 + \frac{4\pi e^2}{\varepsilon_2} \frac{\partial n}{\partial \mu_0}\right]\tilde \varphi_2(k,z,z')= - \frac{4\pi e}{\varepsilon_2} \delta(z-z'), \label{eq13}\\
&\frac{\partial^2\tilde\varphi_3(k,z,z')}{\partial z^2} - k^2 \tilde\varphi_3(k,z,z')=0,\nonumber
\end{align}
\end{widetext}
with the same boundary conditions for $\tilde{\varphi}$ being the same as in \eqref{eq2} for $\varphi$.

For $\tilde{\varphi}_2$, we obtained a linear equation with a variable coefficient, since ${\partial n}/{\partial \mu_0}$ depends on $z$. We note, that the characteristic distance for which ${\partial n}/{\partial \mu_0}$ changes significantly equals $L$. If the expression ${\partial n}/{\partial \mu_0}$ in the equation for $\tilde{\varphi}_2$ is replaced by its average value
\begin{align}
\left<\frac{\partial n}{\partial \mu_0}\right>=\frac{1}{L} \int_0^L \frac{\partial n}{\partial \mu_0} dz = \frac{1}{L} \frac{\partial N}{\partial \mu_0},\nonumber
\end{align}
the equation assumes the form
\begin{align}
\label{eq14}
\frac{\partial^2\tilde\varphi_2(k,z,z')}{\partial z^2} - \tilde k^2 \tilde \varphi_2(k,z,z')= - \frac{4\pi e}{\varepsilon_2} \delta(z-z'), 
\end{align}
where
\begin{align}
\label{eq15}
\tilde{k}=\sqrt{k^2+\frac{4 \pi e^2}{\varepsilon_2 L}\frac{\partial N}{\partial \mu_0}}.
\end{align}

The characteristic distance for which the solution of Eq.~\eqref{eq14} changes significantly is $\tilde{k}^{-1}$. If it is large compared to $L$, i.e., $\tilde{k} L\ll 1$, then $\tilde{\varphi}_2(k, z, z\,')$ is approximately equal to the solution of Eq.~\eqref{eq14} with an average coefficient. Replacing the second equation in the set~\eqref{eq13} with Eq.~\eqref{eq14} and solving the resulting set of equations, we obtain for $\tilde{\varphi}_2(k, z, z\,')$ an expression, formally similar to \eqref{eq5}:
\begin{multline}
\label{eq16}
\tilde \varphi_2(k,z,z') = \frac{2\pi e}{\varepsilon_2 \tilde k} \left\{ e^{-\tilde k|z-z'|} \right. + \\ \left.
\frac{2\tilde \delta}{e^{2\tilde kL} - \tilde \delta^2} [\tilde \delta \cosh{\tilde k(z-z')} + e^{kL} \cosh{\tilde k(z+z'-L)}] \right\}
\end{multline}
where $\tilde{k}$ is defined as in \eqref{eq15} and
\begin{align}
\tilde{\delta}=\frac{\varepsilon\tilde{k}-k}{\varepsilon\tilde{k}+k}. \nonumber
\end{align}
Since $\tilde{k} L\ll 1$ in \eqref{eq16}, we obtain, as it is the case in Eq.~\eqref{eq6}, a ``two-dimensional'' expression
\begin{align}
\label{eq17}
\tilde \varphi(k) = \frac{2\pi e}{\varepsilon_2 \tilde k} \left[\frac{e^{\tilde kL}+\tilde \delta}{e^{\tilde kL}-\tilde \delta}\right].
\end{align}

The inequality $\tilde{k} L\ll 1$ can be satisfied only for the simultaneous fulfillment of the two inequalities
\begin{align}
kL\ll 1 \nonumber
\end{align}
and
\begin{align}
\label{eq18}
\frac{4 \pi e^2{L}}{\varepsilon_2}\frac{\partial N}{\partial \mu_0} \ll 1.
\end{align}
The first inequality means that the expression \eqref{eq17} describes the behavior of the potential $\tilde{\varphi}(\rho)$ for large distances $\rho \gg L$.

Let us elucidate the physical meaning of the inequality \eqref{eq18}. For the approximations assumed in the beginning of the paper [\textit{it follows then that}]
\begin{align}
\mu_0=E_1+k_B T\,\ln\left(e^{\frac{\pi N \hbar^2}{m k_{B} T}}-1\right),\nonumber
\end{align}
where $E_1$ is the energy of the first quantized state [(\textit{i.e., first subband})] and
\begin{align}
\label{eq19}
\frac{\partial N}{\partial \mu_0}=\frac{m}{\pi \hbar^2}\left(1-e^{\frac{\pi N \hbar^2}{m k_{B} T}}\right).
\end{align}
Inserting \eqref{eq19} into \eqref{eq18}, we obtain
\begin{align}
\label{eq20}
\frac{4L}{a}\left(1-e^{\frac{\pi N \hbar^2}{m k_{B} T}}\right)\ll 1,
\end{align}
where $a=\varepsilon_2\hbar^2/me^2$ is the radius of the first  Bohr orbit in the dielectric environment with permittivity $\varepsilon_2$. For the [\textit{two}] limiting cases of a Fermi distribution of electrons in the first subband ${\pi N \hbar^2}/(m k_{B} T)\gg 1$ and for the non-degenerate Boltzmann distribution ${\pi N \hbar^2}/(m k_{B} T)\ll 1$, the inequality~\eqref{eq20} assumes the form
\begin{align}
\label{eq21}
\frac{4L}{a}\ll 1,~\frac{4L}{a}{\frac{\pi N \hbar^2}{m k_{B} T}}\ll 1.
\end{align}
To understand the physical meaning of the requirement (20), let us recall the expression for the Debye radius $r_0$ in a bulk crystal:
\begin{equation}
\label{eq22}
r_0^{-1} = \begin{cases}
2\left(\dfrac{3}{\pi}\right)^{1/6} \sqrt{\dfrac{me^2}{\varepsilon_2 \hbar^2}n^{1/3}} ~~~\mbox{(Fermi case)}\\
\sqrt{\dfrac{4\pi n e^2}{\varepsilon_2 k_B T}}~~~\mbox{(Boltzmann case)}.
\end{cases}
\end{equation}
Let us [\textit{further}] consider the ratio $\alpha={L}/{r_0}$, taking into account that $n={N}/{L}$.
\begin{equation}
\label{eq23}
\alpha = \frac{L}{r_0} = \begin{cases}
2\left(\dfrac{3}{\pi}\right)^{1/6} ( NL^2)^{1/6} \left(\dfrac{4L}{a}\right)^{1/2}~~~\mbox{(Fermi case)}\\
\sqrt{\dfrac{4\pi N \hbar^2}{m k_B T}}\left(\dfrac{4L}{a}\right)^{1/2}~~~\mbox{(Boltzmann case)}
\end{cases}
\end{equation}

Comparing the  expressions \eqref{eq23} for the limiting cases with the left hand sides of the inequalities \eqref{eq21}, we see, that the requirement \eqref{eq20}, which provides the ``two-dimensionality'' of the screening potential for large distances, coincides with the requirement $\alpha \ll 1$. It means, that the film thickness must be smaller than the Debye screening radius in the [\textit{corresponding}] bulk crystal.

Similar to the previous section, let us obtain the expressions of the two-dimensional screening potential $\tilde{\varphi}(\rho)$ for the three limiting cases of the relation between $\varepsilon_1$ and $\varepsilon_2$. We will consider the case of degeneracy, when
\begin{align}
\label{eq24}
\tilde{k}=\sqrt{k^2+\frac{4}{aL}},
\end{align}
keeping in mind, that according to \eqref{eq21}, it is easy to proceed to the non-degenerate case by replacing $a$ with $a{m k_{B} T}/{(\pi N \hbar^2)}$.

\textbf{1.} $\varepsilon \gg 1$. In this case, $1+\tilde{\delta}=2,~1-\tilde{\delta}={2k}/{\varepsilon\tilde{k}}$ and we obtain from \eqref{eq17} (same result was obtained by using the diagram technique in N.S. Rytova, Proc. USSR Academy of Sciences {\bf 163}, 1118 (1965)):
\begin{align}
\label{eq25}
\tilde{\varphi}(k)=\frac{4\pi e}{\varepsilon_2 L\left[k\left(k+\frac{2}{\varepsilon L}\right)+\frac{4}{aL}\right]}.
\end{align}
If the following also holds in that case
\[
\frac{1}{\varepsilon^2}\ll\frac{4L}{a},
\]
then the linear term can be neglected in the denominator of Eq. ~\eqref{eq25}, [\textit{with the result}]
\begin{align}
\tilde{\varphi}(k)=\frac{4\pi e}{\varepsilon_2 L\left(k^2+\frac{4}{aL}\right)}.\nonumber
\end{align}
Then, we obtain for $\rho\gg L$
\begin{align}
\label{eq26}
\tilde\varphi(\rho) = \frac{2e}{{\varepsilon_2} L} K_0 \left( \frac{2\rho}{\sqrt{La}}\right),
\end{align}
where $K_0(x)$ is the zero-order Macdonald function [\textit{modified Bessel function of the second kind}]. For distances $\rho > \rho_0=\frac{1}{2}\sqrt{L a}$, the equation \eqref{eq26} assumes the form
\[
\tilde\varphi(\rho) = \frac{2e}{{\varepsilon_2} L} {\sqrt{\frac{2\rho_0}{\pi \rho}}} e^{-\rho/\rho_0},
\]
from which it follows, that the screened potential is exponentially small beyond the circle with radius $\rho_0$, where
\[
\rho_0 = \begin{cases}
\dfrac{1}{2} \sqrt{La}~~~(\mbox{Fermi case})\\
\dfrac{1}{2} \sqrt{La \dfrac{mk_B T}{\pi N\hbar^2}} = \sqrt{\dfrac{L\varepsilon_2 k_B T}{4\pi N e^2}}~~~(\mbox{Boltzmann case}).
\end{cases}
\]

In case of the Boltzmann distribution, the two-dimensional screening radius $\rho_0$ has the same structure as the three-dimensional one (c.f. Eq.~\eqref{eq22}). For the Fermi distribution, $\rho_0$ does not depend on concentration as a fundamental consequence of [\textit{the Fermi}] statistics in two dimensions.

\textbf{2.} $\varepsilon \approx 1$. Inserting $1+\tilde{\delta}={2\tilde{k}}/({k+\tilde{k}})$ and $1-\tilde{\delta}={2{k}}/({k+\tilde{k}})$ into \eqref{eq17}, we obtain
\begin{align}
\label{eq27}
\tilde \varphi(k) = \frac{2\pi e}{\varepsilon_2 (k+2/a)}
\end{align}
and it follows for $\rho \gg L$
\begin{align}
\label{eq28}
\tilde \varphi(\rho) = \frac{e}{\varepsilon_2} \left\{ \frac{1}{\rho} -\frac{\pi}{a} \left[ \mathbf H_0\left(\frac{2\rho}{a}\right) - N_0 \left(\frac{2\rho}{a}\right)\right]\right\}.
\end{align}
For large distances where $\rho \gg a/2$, Eq. \eqref{eq28} assumes the form
\begin{equation}
\label{eq29}
\tilde \varphi(\rho) = 
\begin{cases}
\dfrac{e a^2}{4\varepsilon_2\rho^3}~~~(\mbox{Fermi case})\\
\dfrac{e a^2}{4\varepsilon_2\rho^3} \left(\dfrac{mk_B T}{\pi N\hbar^2}\right)^2~~~(\mbox{Boltzmann case}).
\end{cases}
\end{equation}
In this case, the screening radius does not exist. At large distances, the potential decays as the inverse cube of the distance.\footnote{Note added in translation: This result has been derived about the same time independently in F. Stern, Phys. Rev. Lett. {\bf 18}, 546 (1967); F. Stern, W. E. Howard, Phys. Rev.  {\bf 163}, 816 (1967).}

\textbf{3.} $\varepsilon \gg 1$. Inserting $1+\tilde{\delta}={2\varepsilon\tilde{k}}/({\varepsilon\tilde{k}+k})$ and $1-\tilde{\delta}={2{k}}/({\varepsilon\tilde{k}+k})$ into \eqref{eq17} and excluding the $k$-independent term (since it does not contribute to the potential at large distances), we obtain
\begin{align}
\label{eq30}
\tilde \varphi(k) = \frac{2\pi e \varepsilon}{\varepsilon_2\left(k+\frac{2\varepsilon}{a}\right)} = \frac{2\pi e}{\varepsilon_1\left(k+\frac{2me^2}{\varepsilon_1 \hbar^2}\right)}. 
\end{align}
The result is equivalent to the expression \eqref{eq27} for the case \textbf{2}, with the replacement of $\varepsilon_2$ by $\varepsilon_1$. Equations~\eqref{eq28} and \eqref{eq29} also hold for the same substitution.

The expressions for the screened potential in cases \textbf{2} and \textbf{3} neither depend on the film thickness $L$ nor on its permittivity $\varepsilon_2$ (in case \textbf{2}, $\varepsilon_2=\varepsilon_1$) and can thus be obtained by passing to the limit of a neutral conducting plane (see below). The expression \eqref{eq26}, applicable for the case \textbf{1}, contains the finite film thickness $L$ and its permittivity $\varepsilon_2$. In this case, there is fundamentally no passage to the limit of the conducting plane from the film of a finite thickness.

\subsection*{Screened potential of a point charge in a conducting plane\footnote{This part was typeset in a small print in the original.}}

{\small The plane $z=0$ is neutral and contains free charges with an average surface concentration $N$. The permittivity of the environment is denoted by $\varepsilon_1$. Let us calculate the potential, created in the $z=0$ plane by a point charge $+e$ placed at the origin of the coordinates.}

{\small In the regions 1 ($z<0$) and 2 ($z>0$), the potential $\varphi(r)$ satisfies the Laplace equation 
\begin{align}
\label{eq31}
\nabla^2\varphi=0.
\end{align}
The potential is bounded at the infinity and the following conditions hold at the $z=0$ boundary
\begin{align}
\label{eq32}
\varphi_1 &= \varphi_2, \nonumber~\\~\\
\frac{\partial \varphi_1}{\partial z} - \frac{\partial \varphi_2}{\partial z} &= \frac{4\pi}{\varepsilon_1} \sigma(\rho),\nonumber
\end{align}
where $\sigma(\rho)$ is the surface charge density in the $z=0$ plane, composed from the point charge at the coordinates origin and the charge induced by the rearrangement of the free electrons in the field of the point charge,
\begin{align}
\sigma(\rho)=e\delta(\vec{\rho})-e\Delta N(\rho). \nonumber
\end{align}}

{\small Let us expand $\varphi({\vec{r}})$ into a double-Fourier-integral:
\begin{align}
\varphi(\vec{r}) =  \int \frac{d\vec{k}}{(2\pi)^2} e^{i \vec{k}\vec{\rho}} \varphi(k,z) \nonumber
\end{align}
From (31) and (32), we obtain for the Fourier components the following equation
\begin{align}
\label{eq33}
\frac{\partial^2 \varphi}{\partial z^2} - k^2\varphi=0,
\end{align}
with the boundary conditions
\begin{equation}
\label{eq34}
\mbox{at}~z=0\quad \begin{cases}
\varphi_1 = \varphi_2\\~\\
\dfrac{\partial \varphi_1}{\partial z} - \dfrac{\partial \varphi_2}{\partial z}= \dfrac{4\pi e}{\varepsilon_1}(1-\Delta N(k)),
\end{cases}
\end{equation}
and the requirement to be bounded to zero at infinity.}

{\small Inserting the solutions of the equation \eqref{eq33} $\varphi_1(k, z)=c_1 e^{kz}$ and $\varphi_2(k, z)=c_2 e^{-kz}$ into the boundary conditions \eqref{eq34} we obtain an equation for the Fourier component of the potential in the $z=0$ plane
\begin{align}
\label{eq35}
\tilde \varphi(k) \equiv \varphi(k,z=0) = \frac{2\pi e}{\varepsilon_1 k} [1-\Delta N(k)].
\end{align}
At large distances, i.e., for small $k$, the quantity $\Delta N$ is linear in $\tilde{\varphi}$
\begin{align}
\label{eq36}
\Delta N = \frac{\partial N}{\partial \mu_0}e\tilde{\varphi}(k).
\end{align}
Inserting \eqref{eq36} into \eqref{eq35}, we find
\begin{align}
\tilde \varphi(k) = \frac{2\pi e}{\varepsilon_1 \left( k+ \frac{2\pi e^2}{\varepsilon_1} \frac{\partial N}{\partial \mu_0}\right)} \nonumber
\end{align}
This result is equivalent to the results \eqref{eq27} and \eqref{eq30} for a film with a finite thickness.
}

~\\
~\\
The author would like to thank V. L. Bonch-Bruyevich for helpful comments.

\end{document}